\documentclass[prd,eqsecnum,nofootinbib,preprint,floatfix]{revtex4}
\usepackage{bm}
\usepackage{multirow}
\usepackage{enumerate}
\reversemarginpar
\def\Tr{{\rm Tr}}

\def\Eq#1{Eq.~(\ref{#1})}

\def\tilde{\widetilde}

\def\chipt{$\chi$PT}
\def\Nsite{N_{\rm site}}

\begin{document}

\vspace*{0.5in}

\title{
On the consistency of the Aoki-phase
\vspace*{0.25in}}

\author{Stephen R. Sharpe\\
\vspace*{0.25in}}

\affiliation{Physics Department, University of Washington, 
Seattle, WA 98195-1560, USA
\vspace*{0.6in}
}

\begin{abstract}
Lattice QCD with two flavors of Wilson fermions can
exhibit spontaneous breaking of flavor and parity,
with the resulting ``Aoki phase'' characterized by the
non-zero expectation value 
$\langle \bar\psi \gamma_5 \tau_3 \psi\rangle\ne0$.
This phenomenon can be understood using the
chiral effective theory appropriate to the Symanzik effective
action. Within this standard analysis, 
the flavor-singlet pseudoscalar expectation value vanishes:
$\langle i \bar\psi \gamma_5 \psi\rangle=0$.
A recent reanalysis has questioned this understanding, arguing 
that either the Aoki-phase is unphysical,
or that there are additional phases in which
$\langle i \bar\psi \gamma_5 \psi\rangle\ne0$. 
The reanalysis uses the properties of probability distribution 
functions for observables built of fermion fields and expansions
in terms of the eigenvalues of the hermitian Wilson-Dirac operator.
Here I show that the standard understanding of
the Aoki-phase can, in fact, be consistent
with the approach used in the reanalysis.
Furthermore, if one assumes that the standard understanding is correct,
one can use the methods of the reanalysis to
derive lattice generalizations of the continuum sum rules of
Leutwyler and Smilga.
\end{abstract}

\maketitle

\section{Introduction}
\label{intro}
New patterns of spontaneous symmetry breaking 
can arise in lattice theories away from the continuum limit.
This paper concerns the example of lattice QCD (LQCD) 
with two flavors of (possibly improved) Wilson fermions. 
Aoki proposed long ago that the apparent
masslessness of the pions at non-zero lattice spacing could
be understood if there is a phase in which parity and flavor
are spontaneously broken~\cite{Aoki1}. 
Numerical evidence for such a phase 
(largely in quenched studies)~\cite{Aoki2} was
subsequently supported by theoretical analyses based on 
the linear sigma-model~\cite{Creutz} and on applying
chiral perturbation theory (\chipt) to the continuum
effective Lagrangian of Symanzik~\cite{ShSi}.
The latter analysis incorporates discretization errors in a systematic
and theoretically well-established way, and uses the standard methods of
continuum \chipt.
Near the continuum limit,
it predicts two possible scenarios, depending
on the sign of an unknown low-energy coefficient.
In one scenario, flavor and parity are broken, and there is an Aoki-phase,
while in the other (the ``first-order'' scenario) there is no
spontaneous breaking of lattice symmetries.
The present paper concerns the former scenario, and thus
assumes that there are choices of gauge and fermion actions which
lead to the appropriate sign of the low-energy coefficient.\footnote{%
Recent unquenched simulations with unimproved twisted-mass
fermions in fact indicate that
the first-order scenario applies for both the
Wilson gauge action~\cite{first-order-Wilson} 
and the tree-level Symanzik improved gauge action~\cite{first-order-evidence},
although an Aoki-phase is observed with the Wilson gauge action 
at stronger coupling~\cite{Aoki-evidence}.
}

This analysis of the Aoki-phase
has been recently questioned by Azcoiti, Di Carlo and Vaquero (ADV)~\cite{ADV},
who study the pattern of
spontaneous symmetry breaking (SSB) using probability distribution
functions (p.d.f.s) of fermionic bilinears~\cite{pdf_fermion}.
Given certain assumptions,
they argue that either the pattern of non-vanishing
condensates differs from that in the usual Aoki-phase 
(and that hermiticity is violated),
or that there are phases additional
to the Aoki-phase in which there is a differing pattern of condensates.
Both possibilities are
in contradiction with the standard \chipt-based analysis.
The \chipt\ analysis (recapped below) predicts a non-zero value only
for the condensate $\langle i\bar\psi\gamma_5\tau_3\psi\rangle$,
and predicts only a single symmetry-breaking phase.

The main purpose of this paper is to point out a loophole in the
assumptions made by ADV, 
one that allows the standard Aoki-phase to be present.
There is no need for the presence of more exotic phases.
Furthermore, if one assumes that the standard 
analysis is correct, one can derive
sum-rules for the eigenvalues of the hermitian Wilson-Dirac operator
on the lattice, analogous to those derived by Leutwyler and Smilga
in the continuum~\cite{LeutSm}.

It will be useful to have in mind the main players in the following 
discussion. These are the zero-momentum
parity and parity-flavor violating bilinears:
\begin{equation}
C_0 = \frac1{\Nsite} \sum_n i \bar\psi \gamma_5 \psi(n)\,,
\qquad {\rm and}\qquad
C_3 = \frac1{\Nsite} \sum_n i \bar\psi \gamma_5 \tau_3 \psi(n)\,.
\label{eq:C0C3}
\end{equation}
Here $n$ labels lattice sites, ${\Nsite}$ is the number of such sites,
and the fields are bare lattice flavor doublets.
Note that $C_0$ and $C_3$ are dimensionless lattice quantities.

The remainder of this paper is organized as follows.
The next section provides a brief summary of the 
\chipt\ argument of Ref.~\cite{ShSi}, and gives some additional
results for the ``$\epsilon$-regime'' that will be needed later.
Section~\ref{sec:pdf} then gives a brief summary of the p.d.f.
method and the results relevant for its application to the Aoki-phase.
An important point is that the results from the p.d.f. analysis
are completely consistent with the predictions
of \chipt, so that what is new about the approach of ADV is their
use of eigenvalue decompositions.
In Sec.~\ref{sec:ADV} I summarize the argument of ADV, pointing
out its inconsistency with the \chipt\ analysis.
The core of this paper is Sec.~\ref{sec:antiADV}, 
in which I explain the loophole in the argument of ADV,
and give two examples of how this loophole might apply.
I also present the above-mentioned sum-rules.
I conclude with a summary and a brief discussion of
generalizations in Sec.~\ref{sec:conc}. 
I include two appendices, the
first describing the derivation of a
result used in Sec.~\ref{sec:antiADV},
the second providing an alternative formulation
of the sum-rules using microscopic spectral densities.

\section{Review of $\chi$PT analysis of Aoki-phase}
\label{sec:aoki}

In this section I recall
the essential features of the analysis of Ref.~\cite{ShSi}.
Vacuum alignment for two flavors of Wilson fermions with no
twisted mass term is determined at leading order
in \chipt\ by minimizing the potential\footnote{%
The analysis in this section is in an effective continuum theory, 
so that fields, masses etc. have their usual continuum dimensions.}
\begin{equation}
{{\cal V}^\chi}(\Sigma) = - \frac{c_1}{4} \Tr(\Sigma + \Sigma^\dagger)
+ \frac{c_2}{16} \left[\Tr(\Sigma+\Sigma^\dagger)\right]^2
\,.
\label{eq:Vchi}
\end{equation}
Here $\Sigma$ is an $SU(2)$ matrix proportional to the
quark condensate, and the coefficients have magnitudes
$c_1\sim m \Lambda_{\rm QCD}^3$ and $c_2\sim a^2 \Lambda_{\rm QCD}^6$,
with $m$ the physical quark mass and $a$ the lattice spacing.\footnote{%
More explicitly, in terms of the low-energy coefficients
in the chiral Lagrangian, 
$c_1 =  2 f^2 B_0 m$ and
$c_2 = - 16 W' W_0^2 a^2$,
where I use the notation of Ref.~\cite{ShWu}.
The physical quark mass is $m= Z_m(m_0-m_c)/a$, 
with $m_0$ the bare Wilson quark mass and $m_c$ the critical mass.}
To produce an Aoki-phase, the
two terms in the potential must have comparable magnitudes,
which occurs if $m\sim a^2\Lambda_{\rm QCD}^3$.
This, together with $m\ll \Lambda_{\rm QCD}$,
is the power-counting used in this paper.
Note that discretization errors linear in $a$ have been
be absorbed into the critical mass $m_c$. 

The Aoki-phase scenario occurs when $c_2 > 0$.
Recalling that $c_1$ is proportional to $m$ (with a positive coefficient
of proportionality), one finds that for $|m|$ large enough that
$|c_1| \ge 2 c_2$, the condensate is aligned with $m$
just as in the continuum:
$\Sigma_0 \equiv \langle \Sigma\rangle = {\rm sign}(m) {\bf 1}$.
For smaller values of $|m|$, however, the potential is minimized by 
$\Sigma_0 = \exp(i \theta_0 \hat n \cdot \vec\tau)$
with $\cos\theta_0= c_1/(2 c_2)$ and $\hat n$ a unit vector.
The $SU(2)$ flavor symmetry $\Sigma\to U \Sigma U^\dagger$ is
then broken to the $U(1)$ subgroup with 
$U=\exp(i\phi \hat n\cdot\vec\tau)$.
This results in two massless Goldstone pions, with
the third pion having a mass proportional to $a$.
The direction of the condensate can be fixed by adding a source term
to the action, and the standard choice is to add the canonical 
twisted-mass term $\mu i\bar\psi\gamma_5\tau_3\psi$, with
$\psi$ the bare lattice $SU(2)$ doublet.
This gives a mass to the (now pseudo-)Goldstone pions proportional
to $\sqrt{\mu}$.
One then sends the volume $V$ to infinity, followed by $\mu\to0$,
so that all pions are massive except in the final limit.
This results in the condensate being aligned
as $\Sigma_0= \cos\theta_0 + i \sin\theta_0 \tau_3$,
with the charged pions being massless and the neutral pion 
having $m_{\pi^0}\sim a \Lambda_{\rm QCD}^2$.\footnote{%
This analysis of course receives corrections from higher order
terms in \chipt, as described in Refs.~\cite{observations,aokibar}.
}

An important point for the subsequent discussion is that
taking $V\to \infty$ before $\mu\to 0$ implies that
$m_{\pi^\pm} L\to\infty$, so that
one is in the so-called ``$p$-regime'' for the charged pions.
Were one to take the limits in the other order,
i.e. $\mu\to0$ and then $V\to\infty$, then the zero-momentum
fluctuations in
the charged pion directions would be unsuppressed, and
one would be in the ``$\epsilon$-regime'' for these 
modes~\cite{epsilonregime}.
I will distinguish between expectation values obtained in these
two orders of limits using the subscripts ``$p$'' and ``$\epsilon$''.
While this has the advantage of linking the limits to familiar names,
one should also keep in mind that
both $p$- and $\epsilon$-regimes are defined more generally.
In particular, they are defined
also for finite $V$---as the regimes in which
$m_\pi L\gg1$ or $\ll 1$, respectively, with $L$ the box size.
I am not making use of the full extent of these regimes,
but rather only the single points in each regime
reached in the limits described above.
Note that $m_{\pi^0} L\sim a L\to\infty$ in both regimes, since $a$
is held fixed,
so that fluctuations in the neutral pion direction are always suppressed.
This means that both regimes for the lattice theory differ
from the corresponding regimes in the continuum.
For example, 
the continuum $\epsilon$-regime has unsuppressed zero-momentum modes
in all three pion directions.

Using the leading order relations between quark bilinears and $\Sigma$,
it follows that, in the Aoki-phase, the parity and flavor-breaking
condensates 
are 
\begin{eqnarray}
\frac{Z_{P}}{a^3} \langle C_3 \rangle_{p} &=& 
i (f^2 B_0/2) \Tr\left(\tau_3\left[\Sigma_0-\Sigma_0^\dagger\right]\right)
= - \sin\theta_0\, 2 f^2 B_0\,,
\label{eq:P3aoki}
\\
\frac{Z_{P_0}}{a^3} \langle C_0 \rangle_{p} &=& 
(f^2 B_0/2) i\Tr\left(\Sigma_0-\Sigma_0^\dagger\right) 
=0 \,.
\label{eq:P0aoki}
\end{eqnarray}
The matching factors and powers of $a$
are needed to convert from lattice to continuum normalization.
Here $f$ and $B_0$ are the standard low-energy coefficients of \chipt,
in terms of which the continuum condensate with a standard mass term
is
\begin{equation}
\langle \bar\psi \psi \rangle_{\rm cont} =
\langle \bar u u +  \bar d d\rangle_{\rm cont}
= - {\rm sign}(m) 2 f^2 B_0
\,.
\end{equation}
On the lattice, only the pseudoscalar condensates can
be determined, due to the mixing of $\bar\psi\psi$ with the identity
operator. Thus it is the results of Eqs.~(\ref{eq:P3aoki})
and (\ref{eq:P0aoki}) that are pertinent. 
I emphasize two features of these results:
(i) The magnitude of the flavor-non-singlet condensate varies as one
moves across the Aoki-phase, with the maximum at $m=c_1=0$
being equal to $|\langle\bar\psi\psi\rangle_{\rm cont}|$ when
appropriately normalized;
(ii) The vanishing of the flavor-singlet pseudoscalar
condensate is true for any $\Sigma_0\in SU(2)$, and is not
special to the particular vacuum of the Aoki-phase. 

This completes the discussion of the analysis of Ref.~\cite{ShSi}.
For the following, it will be necessary to generalize the results
in two ways. The first is to determine the zero-momentum two-point
functions of the bilinears in the $p$-regime.
The results are simple (and can be written in a way
that avoids $Z$-factors):
\begin{eqnarray}
\left\langle
C_3^2 \right\rangle_{p} 
&=& 
\left\langle C_3 \right\rangle_{p}^2 \,,
\label{eq:C3sq_p}
\\
\left\langle C_0^2 \right\rangle_{p} 
&=& 
\left\langle C_0 \right\rangle_{p}^2 = 0 \,.
\label{eq:C0sq_p}
\end{eqnarray}
In words, only the disconnected contributions remain when $\Nsite\to\infty$.
This is because, in the $p$-regime, the connected parts receive
contributions from massive intermediate states, and thus are localized
in space. Although one of the $1/\Nsite$ factors is canceled by
translations, the other remains and causes the contributions to vanish
when $\Nsite\to\infty$.

The second generalization is to the $\epsilon$-regime.
As discussed above, 
zero-momentum fluctuations of $\Sigma$ in the charged-pion directions 
are unsuppressed, and one must integrate the zero-modes over
the Goldstone manifold~\cite{epsilonregime}.
Non-zero-momentum modes can be ignored at leading order.
The result is that the flavor-parity breaking condensate now vanishes
\begin{equation}
\langle C_3 \rangle_{\epsilon} = \frac{a^3}{Z_P}
\int_{\Sigma_0\in SU(2)/U(1)} i (f^2 B_0/2)
\Tr\left(\tau_3\left[\Sigma_0-\Sigma_0^\dagger\right]\right)
= 0\,.
\end{equation}
By contrast, the average of the square of the zero-momentum mode
does not vanish:
\begin{eqnarray}
\langle C_3^2 \rangle_{\epsilon} &=&
\left(\frac{a^3}{Z_P}\right)^2
\int_{\Sigma_0\in SU(2)/U(1)} (i f^2 B_0/2)^2
\left\{
\Tr\left(\tau_3\left[\Sigma_0-\Sigma_0^\dagger\right]\right)
\right\}^2 
\nonumber
\\
&=& 
\frac13 
\left(\frac{a^3}{Z_P}\right)^2
(\sin\theta_0 2 f^2 B_0)^2
\\
&=& 
\frac13 \langle  C_3 \rangle_{p}^2
\,.
\label{eq:c3sq_eps}
\end{eqnarray}
Comparing the final expression to \Eq{eq:C3sq_p}
one sees how averaging over the Goldstone manifold reduces
the size of the squared condensate by the geometrical factor of $1/3$.
For the flavor-singlet pseudoscalar, however, the corresponding
expectation values vanish,
\begin{equation}
\langle C_0 \rangle_{\epsilon} 
=
\langle C_0^2 \rangle_{\epsilon} 
=
0
\,,
\label{eq:c0sq_eps}
\end{equation}
because the vacuum manifold is simply the origin.

It is straightforward to
extend these results to higher powers of the $C_a$,
with results that are quoted and used in Appendix~\ref{app:alternative}.

\section{Probability density functions for fermion bilinears}
\label{sec:pdf}

The ADV analysis of SSB for Wilson fermions uses 
probability distribution functions for fermion bilinears.
Here I briefly recall the essential properties of these p.d.f.s.
The p.d.f. is familiar for scalar field theories, where it
is defined by inserting 
$\delta(C-\frac1{\Nsite} \sum_n \phi_n)$ in the functional integral
(considering here a real scalar field).
It is related to the constraint effective potential,
$P(C)=\exp[-{\Nsite} {\cal V}_{\rm constr.}(C)]$ 
and is a useful tool for studying symmetry breaking.
In particular, in the absence of source terms, it is invariant
under the symmetries of the action. 
For example, if there is a $\phi\to-\phi$ symmetry,
and this is spontaneously broken, then
$P(C) = (1/2)[\delta(C-C_0) + \delta(C+C_0)]$ when ${\Nsite}\to\infty$, with
$C_0$ the magnitude of the expectation value in the presence of an
infinitesimal source. If the symmetry is unbroken, then
$P(C) = \delta(C)$. 

The construction of a p.d.f. is generalized to fermion bilinears
in Ref.~\cite{pdf_fermion}.
It is a non-trivial result that the resulting p.d.f. can be used in the
same way as for scalar field theories, and, in particular,
as a tool to study SSB.
In the present instance the p.d.f.s of interest are 
those for $C_{0}$ and $C_3$.
If one could calculate $P(C_0)$ and $P(C_3)$ using the
method of Ref.~\cite{pdf_fermion} then one could
deduce whether SSB occurs and the nature of any broken phases.

ADV make particular use of $P(C_0)$ and $P(C_3)$ in the ``Gibbs state'',
which means here that one evaluates 
them with no twisted-mass source term,
but takes the ${\Nsite}\to \infty$ limit. 
As noted in the previous section, this puts the theory in
the $\epsilon$-regime, in which symmetries are manifest.
Thus the expectation values of $C_{a}$, $a=0,3$ vanish,
because of parity and flavor-parity respectively:
\begin{equation}
\langle C_{a}\rangle_{\rm \epsilon} \equiv
\lim_{N_{\rm site}\to\infty} \int dC_{a}\ P(C_{a})\ C_{a} =0\,.
\label{eq:Capdf}
\end{equation}
Note that this result gives no information about SSB,
since it holds
irrespective of whether the symmetries would
spontaneously break were the limits taken in the other
order (${\Nsite}\to\infty$ followed by the twisted-mass source term
vanishing).
It is also possible, as suggested by ADV, 
that there are several vacua, each with differing patterns of
condensates, which are averaged over in the Gibbs state.
leading to a vanishing result.

Higher moments of the p.d.f.s are, however, order parameters for
SSB. Following ADV, I 
focus mostly on the second moment, 
for which one expects~\cite{pdf_fermion,ADV}
\begin{equation}
\langle C_a^2\rangle_{\epsilon} = 
\lim_{{\Nsite}\to\infty} \int dC_{a}\ P(C_{a})\ C_{a}^2 =
 f_{\rm geom} \times |\langle C_a \rangle_{p}|^2\,.
\label{eq:Gibbs}
\end{equation}
Here $f_{\rm geom}$ is a non-vanishing
geometrical factor which depends on the vacuum manifold.
It follows from (\ref{eq:Gibbs}) that, if there
is no SSB, and $\langle C_a\rangle_p=0$, then
$\langle C_a^2\rangle_\epsilon$ will vanish.
On the other hand, if there is SSB and $\langle C_a\rangle_p\ne0$, 
then $\langle C_a^2\rangle_\epsilon$ will be non-vanishing.

The value of $f_{\rm geom}$ can be determined if the vacuum manifold
is known. For a single vacuum, $f_{\rm geom}=1$,
while for the continuous $U(1)$ (complex scalar field), 
$SU(2)/U(1)$ (Aoki-phase)
and $SU(2)$ (continuum chiral symmetry breaking)  manifolds
the factor is $f_{\rm geom}=1/2$, $1/3$ and $1/4$, respectively.
These are simply obtained by averaging the squared projection of
the field onto a fixed axis over the respective manifolds.
More complicated vacuum manifolds with disconnected
components would lead to other, less simple, values of $f_{\rm geom}$.

ADV apply this methodology to the Aoki-phase.
If the standard analysis holds, then
$\langle C_0 \rangle_{p}=0$ while 
$\langle C_3 \rangle_{p}\ne 0$,
and one then finds
\begin{eqnarray}
\langle C_3^2\rangle_{\epsilon} &=& 
\frac{|\langle C_3 \rangle_{p}|^2}{3}\ne 0\,.
\label{eq:c3sq_pdf}
\\
\langle C_0^2\rangle_{\epsilon} &=& 0\,,
\label{eq:c0sq_pdf}
\end{eqnarray}
The issue in the following is whether these
results are correct, and in particular, whether they are 
consistent with expressions in terms of eigenvalues of
the hermitian Wilson-Dirac operator.
If not, the standard understanding of the Aoki-phase must be wrong.
This holds also for the extension of the above results
to higher powers of the $C_a$, which are straightforward to derive,
and which are quoted and used in Appendix~\ref{app:alternative}.

I close this section by noting that the results
of the p.d.f. analysis can also be obtained using \chipt.
This is shown by the consistency of 
Eqs.~(\ref{eq:c3sq_eps}) and (\ref{eq:c0sq_eps}) from the previous
section with Eqs.~(\ref{eq:c3sq_pdf}) and (\ref{eq:c0sq_pdf}).
This consistency is, in fact, preordained,
because the required average over the vacuum manifold is
identical in the two approaches.
Each approach has its strengths and weaknesses:
\chipt\ adds the specific prediction
for $\langle C_3 \rangle_{p}$, \Eq{eq:P3aoki},
while the p.d.f. analysis does not require an expansion in
$m$ and $a$.
Still, for small $m\sim a^2\Lambda_{\rm QCD}^3$, 
as considered here, one does not need the p.d.f. methodology
to pose the puzzle noted by ADV.

\section{The argument of ADV}
\label{sec:ADV}

ADV base their argument on the expressions for the expectation values
in terms of the (real dimensionless)
eigenvalues of the hermitian Wilson-Dirac operator,
$H_W=\gamma_5 D_W$. Denoting these
eigenvalues by $\lambda_j$, and including a bare lattice twisted mass,
$\mu_0$, one has
\begin{eqnarray}
\langle C_0 \rangle &=& \frac{2i}{{\Nsite}}
\left\langle \sum_j \frac{\lambda_j}{\mu_0^2+\lambda_j^2}\right\rangle
\ \stackrel{\mu_0\to0}{\longrightarrow}\
\frac{2i}{{\Nsite}}
\left\langle \sum_j \frac{1}{\lambda_j}\right\rangle
\label{eq:C0lam}
\\
\langle C_3 \rangle &=& -\frac{2}{{\Nsite}}
\left\langle \sum_j \frac{\mu_0}{\mu_0^2+\lambda_j^2}\right\rangle
\ \stackrel{\mu_0\to0}{\longrightarrow}\
0
\label{eq:C3lam}
\\
\langle C_{a}^2 \rangle &=&  
\langle C_{a}^2 \rangle_{\rm disc} + \langle C_{a}^2 \rangle_{\rm conn} 
\label{eq:Calam}
\\  
\langle C_0^2 \rangle_{\rm disc} &=&
- \frac{4}{{\Nsite}^2}
\left\langle\left(\sum_j \frac{\lambda_j}
                              {\lambda_j^2+\mu_0^2}\right)^2\right\rangle
\ \stackrel{\mu_0\to0}{\longrightarrow}\ 
- \frac{4}{{\Nsite}^2}
\left\langle\left(\sum_j \frac{1}{\lambda_j}\right)^2\right\rangle
\label{eq:C0sqdisclam}
\\
\langle C_0^2 \rangle_{\rm conn} &=&
\langle C_3^2 \rangle_{\rm conn} =
\frac{2}{{\Nsite}^2}
\left\langle \sum_j \frac{\lambda_j^2-\mu_0^2}
                         {(\lambda_j^2+\mu_0^2)^2}\right\rangle
\ \stackrel{\mu_0\to0}{\longrightarrow}\
\frac{2}{{\Nsite}^2}
\left\langle \sum_j \frac{1}{\lambda_j^2}\right\rangle
\label{eq:Csqconnlam}
\\
\langle C_3^2 \rangle_{\rm disc} &=&
\frac{4}{{\Nsite}^2}
\left\langle\left(\sum_j \frac{\mu_0}{\lambda_j^2+\mu_0^2}\right)^2\right\rangle
\ \stackrel{\mu_0\to0}{\longrightarrow}\ 
0
\label{eq:C3sqdisclam}
\end{eqnarray}
The first result on each line can be used to obtain the
``$p$-regime'' expectation values (i.e. ${\Nsite}\to\infty$ and
then $\mu_0\to0$). Taking the limits in this order means that
the spectrum of eigenvalues becomes continuous.
The second result on each line (i.e. that with $\mu_0=0$) 
gives, once ${\Nsite}\to\infty$, 
the ``$\epsilon$-regime'' expectation values.

Consider first the linear moments, Eqs.~(\ref{eq:C0lam}) and
(\ref{eq:C3lam}). Applying a parity transformation flips the
sign of $\lambda$ and $\mu_0$, and thus averaging over a configuration
and its parity conjugate leads to a vanishing $\langle C_0\rangle$ 
in both $p$- and $\epsilon$-regimes.\footnote{%
At $a\ne0$ one does not have an index theorem and the corresponding
zero-modes related to topology. There can be isolated zero-modes
on some configurations, as discussed further in the next section,
but these will introduce an additional suppression of $\mu_0^2$
from the fermion determinant, and thus do not contribute when
$\mu_0\to0$ to either of the linear moments. This holds also
for the quadratic moments.}
This is consistent with the standard expectations,
Eqs.~(\ref{eq:P0aoki}) and (\ref{eq:Capdf}).
The prediction of (\ref{eq:C3lam}) that
$\langle C_3\rangle_\epsilon=0$ is also consistent with \Eq{eq:Capdf}.
In the $p$-regime, however, one can obtain a non-vanishing expectation
\begin{equation}
\langle C_3\rangle_p= - 
\lim_{\mu_0\to0}\lim_{{\Nsite}\to\infty}
\frac2{\Nsite} \left\langle \sum_j
\frac{\mu_0}{\mu_0^2+\lambda_j^2}\right\rangle
= -2\pi \langle \rho_U(0)\rangle_p 
\equiv -2\pi \rho(0) 
\,.
\label{eq:C3resolution}
\end{equation}
This is the standard Banks-Casher relation applied to
the present context, 
in which $\rho_U(\lambda)$ is the eigenvalue density of $H_W$
per unit (dimensionless) volume
on a given configuration 
(which can be defined when ${\Nsite}\to\infty$) and
$\rho(\lambda)$ is its average over configurations.
For the Aoki-phase scenario to hold one must have
$\rho(0)\ne 0$, and to match with the \chipt\ prediction
(\ref{eq:P3aoki}) requires
\begin{equation}
\rho(0) 
= \frac{a^3}{Z_P}\frac{\sin\theta_0 f^2 B_0}{\pi}
\label{eq:rho0result}
\,.
\end{equation}
Thus far the results are conventional and uncontroversial.

The apparent problems with the Aoki-phase concern the
expectation value $\langle C_0^2\rangle$. 
In the standard scenario, this is expected
to vanish in both the $\epsilon$-regime [Eqs.~(\ref{eq:c0sq_eps})
and (\ref{eq:c0sq_pdf})]
and $p$-regime [Eq.~(\ref{eq:C0sq_p})].
The expressions in terms of 
eigenvalues---Eqs.~(\ref{eq:Calam}--\ref{eq:Csqconnlam})---do not, 
however, vanish under parity averaging,
reflecting the fact that $C_0^2$ is even under parity.
The issue is whether they vanish when ${\Nsite}\to\infty$.

ADV argue that, assuming the standard properties of
$C_3^2$ in the Aoki-phase, $\langle C_0^2\rangle$ does not
vanish, in contradiction to the standard Aoki-phase scenario. 
They reach this conclusion by considering the
two possibilities for the behavior of the eigenvalues under parity:
\begin{enumerate}
\item
$\rho_U(\lambda)=\rho_U(-\lambda)$:
the eigenvalue distribution for $\mu_0=0$ becomes even in
$\lambda$ on each configuration when ${\Nsite}\to\infty$.
This is what happens if averaging over an infinite
volume effectively includes a parity average,
which is what one would expect.
In this case $\langle C_0^2\rangle_{\rm disc}$ vanishes
in both regimes since the summand in
(\ref{eq:C0sqdisclam}) is odd in $\lambda$.
ADV focus (in their section IV) on the $\epsilon$-regime.
Since $\langle C_3^2\rangle_{{\rm disc},\epsilon}$ vanishes 
identically [see eq.~(\ref{eq:C3sqdisclam})], it follows that
\begin{equation}
\langle C_0^2\rangle_{\epsilon} =
\langle C_0^2\rangle_{{\rm conn,}\epsilon}=
\langle C_3^2\rangle_{{\rm conn,}\epsilon}=
\langle C_3^2\rangle_{\epsilon} \,.
\label{eq:ADV2}
\end{equation}
Thus if $\langle C_3^2\rangle_{\epsilon}$ is non-vanishing,
then so is $\langle C_0^2\rangle_{\epsilon}$.
This is in manifest contradiction with
Eqs.~(\ref{eq:c3sq_eps}) and (\ref{eq:c0sq_eps})
[or equivalently with 
Eqs.~(\ref{eq:c3sq_pdf}) and (\ref{eq:c0sq_pdf})],
and the standard picture of the Aoki-phase fails.
To obtain consistency with (\ref{eq:ADV2})
ADV postulate the presence of an additional phase with
non-vanishing $\langle C_0\rangle_p$.
\item
$\rho_U(\lambda)\ne\rho_U(-\lambda)$:
the eigenvalue distribution is not symmetric on a given configuration,
even in the ${\Nsite}\to\infty$ limit.
In this part of the argument (section III of their paper)
ADV keep $\mu_0$ non-zero (they call it $m_t$)
while sending ${\Nsite}\to\infty$, 
and thus they are working in the $p$-regime.\footnote{%
In general, $\rho_U(\lambda)$ depends on $\mu_0$.
Here, however, $\mu_0$ is infinitesimal, so the flip
in sign of $\mu_0$ under parity does not affect $\rho_U(\lambda)$.}
Since there are then no exactly massless particles,
it follows, as discussed above,
that connected two-point functions vanish when ${\Nsite}\to\infty$.
ADV conclude that 
\begin{equation}
\langle C_0^2\rangle_{p} = \lim_{{\Nsite}\to\infty}
\langle C_0^2\rangle_{\rm disc} = 
-4 \left\langle \left[\int d\lambda \frac{\rho_U(\lambda)}{\lambda}\right]^2
                                     \right\rangle\,,
\label{eq:ADV1}
\end{equation}
which, by assumption, is negative definite.
This is in contradiction with the vanishing result (\ref{eq:C0sq_p})
expected in the Aoki-phase.\footnote{%
ADV expand further on the implications of
$\langle C_0^2\rangle_{p}<0$, which violates hermiticity
in the continuum limit. For my purposes 
it is, however, sufficient to show that the result  
(\protect\ref{eq:ADV1}) is in 
contradiction with the Aoki-phase expectations.}
\end{enumerate}
Having considered both symmetric and asymmetric $\rho_U(\lambda)$,
and finding both in contradiction with the Aoki-phase,
ADV conclude that the standard Aoki-phase scenario must be
incomplete.

I emphasize that, if the argument of ADV is correct,
then one must conclude 
that the standard \chipt-based analysis is {\em incorrect}.
For small enough $m$ and $a^2$, the 
\chipt\ analysis unambiguously predicts only one possible pattern
of symmetry breaking, 
in which $\langle C_0^2\rangle_{p}=\langle C_0^2\rangle_{\epsilon}=0$.
This is simply in contradiction with both of the possibilities
enumerated above, and in particular with Eqs.~(\ref{eq:ADV2}) 
and (\ref{eq:ADV1}).

\section{Consistency of the Aoki-phase with eigenvalue sums}
\label{sec:antiADV}

The conclusion of ADV's argument is surprising.
One would not expect that the Aoki-phase scenario, 
based as it is on a straightforward application of 
the methods of effective field theory, could be invalidated
by simple properties of eigenvalues of $H_W$,
especially since these properties do not appear
to be in conflict with the assumptions of the \chipt\ analysis.

In fact, there is a loophole in the argument of ADV.
This arises because the summand in Eq.~(\ref{eq:C0sqdisclam})
is infrared divergent in the $\epsilon$-regime. 
This allows  an asymmetry in the spectrum which is subleading as
${\Nsite}\to\infty$, and which does not contribute 
in the infrared regulated $p$-regime, to nevertheless contribute 
to $\epsilon$-regime expressions.
In this way it is possible for $\langle C_0^2\rangle_{{\rm disc},p}=0$
while $\langle C_0^2\rangle_{{\rm disc},\epsilon}\ne 0$.
Disconnected and connected contributions
to $\langle C_0^2\rangle_{\epsilon}$, which have opposite signs,
can then cancel, leading to the desired result
$\langle C_0^2\rangle_{\epsilon}=0$.

I stress that showing this possibility exists simply demonstrates
that one cannot make a judgment about the Aoki-phase scenario using
the basic properties of the eigenvalues of $H_W$. 
The Aoki-phase is an allowed option. To determine
whether it actually occurs, however, requires further input,
such as that provided by applying
\chipt\ to the Symanzik effective Lagrangian.
Using this input, which predicts the Aoki-phase (assuming $c_2>0$),
one can turn the ADV argument around and deduce
sum-rules that the eigenvalues must satisfy. These are analogues
of the Leutwyler-Smilga sum-rules for continuum QCD~\cite{LeutSm}.

That the expectation values $\langle C_0^2\rangle_{{\rm disc},p}$
and $\langle C_0^2\rangle_{{\rm disc},\epsilon}$ can differ is 
an example of the non-commutativity of the
$\mu_0\to0$ and ${\Nsite}\to\infty$ limits.
This non-commutativity
is familiar from the properties of $\langle C_3\rangle$ noted above
[see \Eq{eq:C3resolution} and preceding discussion].
Another example is the behavior of
$\langle C_0^2\rangle_{\rm conn}=\langle C_3^2\rangle_{\rm conn}$,
and it will be useful to describe this as a warm-up exercise
before explaining the loophole.

In the $p$-regime one expects
$\langle C_0^2\rangle_{\rm conn}$
to vanish because only massive intermediate states contribute.
To see how this works with eigenvalues, one writes
\begin{equation}
 \langle C_0^2\rangle_{{\rm conn},p}= 
\lim_{\mu_0\to0} \lim_{{\Nsite}\to\infty} \frac2{\Nsite} 
\int d\lambda\ \frac{\lambda^2-\mu_0^2}
                             {(\lambda^2+\mu_0^2)^2}\ \rho(\lambda)
\,.
\end{equation}
Here I have used the result that averaging over configurations
allows one to define a continuous density $\rho(\lambda)$ prior to
taking ${\Nsite}\to\infty$, and I am assuming that $1/{\Nsite}$ and $\mu_0$
are small enough that $\rho(\lambda)$ does not depend on them.
The key point is that the integral on the right-hand-side is finite,
so that $\langle C_0^2\rangle_{{\rm conn}, p}\to0$
when ${\Nsite}\to\infty$ due to the overall factor of $1/{\Nsite}$.
The only possible sources of divergence in the integral are the infrared
and ultraviolet regions.
Only the first two terms
in the Taylor expansion of $\rho$ about $\lambda=0$ can lead to
infrared divergences, but their contributions vanish for
any non-zero $\mu_0$.
The ultraviolet divergence (arising because $\rho\propto \lambda^3$ for
$\lambda\gg a\Lambda_{\rm QCD}$)
is regulated on the lattice because
there is cut-off on eigenvalues, $\lambda_{\rm max}\approx 1$.\footnote{%
That the ultraviolet divergence is subleading in $1/{\Nsite}$
is as in the continuum, as has been discussed in 
Ref.~\cite{LeutSm}. This result applies also for 
the other eigenvalue sums considered below.}

By contrast, in the $\epsilon$-regime, $\langle C_0^2\rangle_{\rm conn}$ 
has the infrared-divergent summand $1/({\Nsite}\lambda_j)^2$.
Since $\rho(0)\ne 0$, the low eigenvalues are approximately
uniformly distributed with spacing 
$\Delta \lambda \sim 1/[{\Nsite}\;\rho(0)]$
and with $\lambda_{\rm min} \sim 1/[{\Nsite}\;\rho(0)]$.
They thus give a non-vanishing contribution to 
$\langle C_0^2\rangle_{\rm conn}$ when ${\Nsite}\to\infty$
[one that cannot be represented as an integral over $\rho(\lambda)$].
This is qualitatively consistent with the expectation
(\ref{eq:c3sq_pdf}) from the p.d.f. analysis. 
To agree quantitatively with the
\chipt\ result (\ref{eq:c3sq_eps}) 
requires that the following sum-rule hold:
\begin{equation}
\lim_{{\Nsite}\to\infty} 
\frac{2}{{\Nsite}^2}
\left\langle \sum_j \frac{1}{\lambda_j^2}\right\rangle
=
\frac{4\pi^2}{3} \rho(0)^2
=
\frac{4}{3} 
\left\{\frac{a^3}{Z_P}{\sin\theta_0 f^2 B_0}\right\}^2
\,.
\label{eq:C3sumrule}
\end{equation}
This constrains the distribution of the small eigenvalues
of $H_W$. Note that it must hold separately for each
value of $m$ throughout the Aoki-phase.

With this warm-up completed I now return to main quantity
of interest, $\langle C_0^2\rangle_{\rm disc}$. For the
Aoki-phase to be consistent this quantity must vanish in
the $p$-regime and cancel $\langle C_0^2\rangle_{\rm conn}$
in the $\epsilon$-regime.
Vanishing in the $p$-regime requires
\begin{equation}
\langle C_0^2\rangle_{{\rm disc},p}
= - \lim_{\mu_0\to0}
\left\langle \left(
\int d\lambda \frac{\lambda[\rho_U(\lambda)-\rho_U(-\lambda)]}
                   {\lambda^2+\mu_0^2}
\right)^2\right\rangle
=0\,.
\end{equation}
Since the integrand is finite in the infrared for any non-zero $\mu_0$,
this relation is satisfied if the density is symmetric:
\begin{equation}
\delta\rho_U(\lambda) \equiv \rho_U(\lambda) - \rho_U(-\lambda) = 0\,.
\label{eq:rhoUsymm}
\end{equation}
I stress that the condition (\ref{eq:rhoUsymm}) concerns only the
spectrum in the ${\Nsite}\to\infty$ limit---indeed,
it is only in this limit that
$\rho_U$ (and thus $\delta\rho_U$) becomes well-defined.

The question then is how $\langle C_0^2\rangle_{{\rm conn},\epsilon}$
can be non-zero and cancel with $\langle C_0^2\rangle_{{\rm disc},\epsilon}$,
i.e. how the sequence of equalities in \Eq{eq:ADV2} of ADV's first 
possibility can fail.
From Eqs.~(\ref{eq:C0sqdisclam}), (\ref{eq:Csqconnlam}) and
(\ref{eq:C3sumrule}) the requirement is that
\begin{equation}
 \lim_{{\Nsite}\to\infty} \frac{1}{{\Nsite}^2}
\left\langle\left(\sum_j \frac{1}{\lambda_j}\right)^2
\right\rangle 
= \frac{\pi^2}3 \rho(0)^2
\,,
\label{eq:C0sumrule}
\end{equation}
This relation must hold throughout the Aoki-phase
(and, in fact, outside this phase too, where $\rho(0)$ vanishes).
The issue is whether the left-hand-side can be non-vanishing given
the symmetry property (\ref{eq:rhoUsymm}).
This is possible if there is an  asymmetry which, 
while vanishing when ${\Nsite}\to\infty$,
is enhanced by the infrared divergence in the summand so
that $\sum_j 1/(N_{\rm site}\lambda_j)$ does {\em not} become 
$\int d\lambda\, \rho_U(\lambda)/\lambda$ when ${\Nsite}\to\infty$.
If it did have this limit then the resulting integral would
vanish given the symmetry of $\rho_U(\lambda)$
and one would be back to the first inconsistency noted by ADV.\footnote{%
The limit in \Eq{eq:Csqconnlam} gives
the principal part, which vanishes if $\rho_U$ is symmetric.}

To show that it is possible for the sum-rule (\ref{eq:C0sumrule}) 
to be satisfied I need to recall the properties of 
the spectrum of $H_W$.
There is a wealth of literature on this topic, and I will use
particularly the results and insights from 
Refs.~\cite{Iwasaki,SmitVink,NN95,Edwards,Adams,GS}.
As noted above, exact zero-modes of $H_W$ are suppressed by
its determinant and are not relevant.
What is important is that the spectrum is
known to be asymmetric on almost all configurations.
This asymmetry is a remnant of the exact zero-modes of
the Dirac operator which
are present in the continuum limit on topologically non-trivial
configurations. On the lattice
these would-be exact zero modes end up as near-zero modes of $H_W$
and lead to an asymmetry, as described in more detail in
Appendix~\ref{app:top}. 
I expect that the typical magnitude of the resulting asymmetry
(defined on a given configuration
as the difference between the number of modes
with $\lambda_j>0$ and $\lambda_j <0$) depends on
$a$ and the lattice volume $V=a^4 N_{\rm site}$ as
\begin{equation}
|n_{\rm asym}| \sim a \sqrt{V} \Lambda_{\rm QCD}^3
\,.
\label{eq:nasym}
\end{equation}
There are several arguments which support this
parametric dependence. The most simple is to note that,
in the continuum, the typical number of zero modes scales as
$|n_{\rm zero}| \sim \sqrt{m V \rho_D(0)}$,
with $\rho_D(0)\sim\Lambda_{\rm QCD}^3$ the 
eigenvalue density (per unit volume) of the Dirac
operator~\cite{LeutSm}.
This result is for the $p$-regime, which is appropriate
since the neutral pion is in its $p$-regime on the lattice.
Now, in the Aoki-phase, symmetry breaking by mass terms
competes with that from $O(a^2)$ discretization effects,
the latter being dominant in the center of the phase.
Thus it is plausible that one can use the continuum formula
with the replacements $m\to a^2\Lambda_{\rm QCD}^3$
and $n_{\rm zero}\to n_{\rm asym}$, leading to \Eq{eq:nasym}.
Further arguments in support of this relation are given
in Appendix~\ref{app:top}.

Another result needed below is
that would-be exact zero-modes are expected to 
have eigenvalues shifted to
$\lambda\sim (a\Lambda_{\rm QCD})^{q}$ by discretization effects.
Here $q$ is an unknown power that I argue below may be $q=3$.

The result (\ref{eq:nasym}) for the spectral asymmetry is consistent
with the $p$-regime requirement (\ref{eq:rhoUsymm}), because one must
divide by $N_{\rm site}$ to obtain the spectral density:
\begin{equation}
\int d\lambda\ \delta\rho_U(\lambda) =
\lim_{N_{\rm site}\to\infty} \frac{n_{\rm asym}}{N_{\rm site}} = 0
\,.
\end{equation}
To satisfy the $\epsilon$-regime requirement (\ref{eq:C0sumrule})
requires consideration of how the asymmetry depends on $\lambda$.
For illustration assume that there are more positive than
negative eigenvalues on a particular configuration. 
Due to eigenvalue repulsion, one expects that 
the extra eigenvalues will impact the spectrum over the region 
$0< \lambda < (a\Lambda_{\rm QCD})^q$.
To satisfy (\ref{eq:C0sumrule}) this impact
must be appropriately peaked at small $\lambda$.
I give two examples of how this could work.
\begin{itemize}
\item
The first is simple: I assume that the
spectral asymmetry manifests itself by the presence of
$O(1)$ extra small (positive) eigenvalues with 
$\lambda_j \sim 1/[N_{\rm site} \rho(0)]$,
while the remaining eigenvalues giving rise to the
asymmetry are distributed in such a way
as to give a contribution to $\sum_j 1/(\Nsite\lambda_j)$
which vanishes when $\Nsite\to\infty$.
As shown by the second example, 
this requires the asymmetry to be less peaked than a
$1/\sqrt\lambda$ singularity.

With these assumptions, $\sum_j 1/(\Nsite\lambda_j)\sim \rho(0)$,
and so the square of this sum gives
$\langle C_0^2\rangle_{{\rm disc},\epsilon} \sim \rho(0)^2$.
This has the correct magnitude to allow the sum-rule
(\ref{eq:C0sumrule}) to be satisfied.

Note that in this example I am not making particular use
of the dependence (\ref{eq:nasym}) of the asymmetry on $a$ and $V$
(nor of the arguments given in Appendix~\ref{app:top}).
It is, however, somewhat artificial to assume an $O(1)$ 
delta-function-like contribution to the asymmetry.
\item
In the second example, the asymmetry is spread continuously
over the range $0<\lambda< (a\Lambda_{\rm QCD})^q$.
For finite ${\Nsite}$, the discrete sum over eigenvalues can be
approximated by an integral aside from end effects which
can be accounted for choosing the limits of integration
appropriately. In this sense, one can consider $\delta\rho_U(\lambda)$
for finite ${\Nsite}$. The behavior I assume is
\begin{equation}
\delta\rho_U(\lambda)
\sim \frac{(a\Lambda_{\rm QCD})^{3-q/2}}{\sqrt{\lambda}\sqrt{\Nsite}}\,,
\end{equation}
in which the key feature is the $1/\sqrt{\lambda}$ singularity.
The other factors are chosen so that one obtains the
desired spectral asymmetry:
\begin{equation}
n_{\rm asym} = \Nsite \int_0^{(a\Lambda_{\rm QCD})^{q}} 
d\lambda\ \delta\rho_U(\lambda)
\sim (a \Lambda_{\rm QCD})^3 \sqrt{\Nsite}
\sim a \sqrt{V} \Lambda_{\rm QCD}^3 
\,.
\end{equation}
The sum in (\ref{eq:C0sumrule}) 
can then be approximated by an integral with
the lower limit regulated with a quantity of order
$\lambda_{\rm min}\sim 1/[\Nsite\;\rho(0)]$:
\begin{equation}
\sum_j \frac1{\Nsite\lambda_j}
\approx
\int_{1/[\Nsite\;\rho(0)]}^{(a\Lambda_{\rm QCD})^{q}} 
d\lambda \frac{\delta\rho_U(\lambda)}
     {\lambda}
\sim
(a\Lambda_{\rm QCD})^{3-q/2} \rho(0)^{1/2}
\sim 
(a\Lambda_{\rm QCD})^{(9-q)/2}
\,.
\label{eq:finalsumresult}
\end{equation}
Here I have used $\rho(0)\sim (a\Lambda_{\rm QCD})^3$
from \Eq{eq:rho0result}.
The key point is that the result 
(\ref{eq:finalsumresult}) has a non-zero (and non-infinite)
limit as $\Nsite\to\infty$. Inserting 
(\ref{eq:finalsumresult}) into \Eq{eq:C0sumrule}
then leads to a non-zero result for
$\langle C_0^2\rangle_{{\rm disc},\epsilon}$.

In addition, recalling that $\rho(0)\propto (a\Lambda_{\rm QCD})^3$,
one sees that for the $a$ dependence on both
sides of Eq.~(\ref{eq:C0sumrule}) to match
requires $q=3$. 

While this example is perhaps more realistic than the
first, I stress that it only works if $\delta\rho$ diverges as
$1/\sqrt\lambda$ and not for other powers.
\end{itemize}

I do not know if either of these examples represents the
actual behavior. Presumably it should be possible
to determine more
detailed information on the distributions of low eigenvalues,
as has been done in the continuum limit using
the methods of random matrix theory.
A small step in this direction has been taken in
Ref.~\cite{SSgap}.

The description of the loophole given above is somewhat awkward
and unsystematic. This shortcoming can be addressed in part by
formulating the required consistency conditions in terms of
the microscopic spectral density and correlations.
This is done in Appendix~\ref{app:alternative}.
Also included in this appendix is some discussion of how
the consistency conditions extend to higher orders
(corresponding to sum-rules with higher overall powers of
$\lambda^{-1}$).

\section{Conclusions}
\label{sec:conc}

The major aim of this paper has been to show the Aoki-phase
scenario is not ruled out by the arguments of ADV.
The examples presented in the previous section
demonstrate this---the set of possibilities considered by ADV is
incomplete.
The ``survival'' of the Aoki-phase 
is consistent with the intuition
that one cannot rule out the results of \chipt\
using only general properties of eigenvalues of $H_W$.

If one accepts the standard analysis of the Aoki-phase,
then one finds non-trivial conditions that must be obeyed by the
eigenvalues of the hermitian Wilson-Dirac operators.
These are the sum-rules (\ref{eq:C3sumrule}) and (\ref{eq:C0sumrule}),
which can also be formulated
as constraints on integrals of microscopic
spectral correlators [Eqs.~(\ref{eq:C3SRA}) and 
(\ref{eq:C0SRA}) respectively].
There are, in fact, an infinite set of these sum-rules, involving
products of any even number of inverse-eigenvalues.\footnote{%
I thank Vicente Azcoiti and collaborators for stressing to me the
importance of the presence of this infinite tower of sum-rules.}
While at first sight it may seem daunting that the eigenvalues of
$H_W$ must be distributed so as to satisfy all these sum-rules,
I show in Appendix~\ref{app:alternative} how each 
sum-rule constrains an essentially independent eigenvalue
correlation function, making it more plausible that they
can all be satisfied. 
The need to satisfy an infinite set of sum-rules is not special
to the lattice theory. Indeed,
for actions with chiral symmetry (as in the
continuum analysis of Leutwyler and Smilga), the eigenvalues
of the Dirac operator in each topological charge sector
must satisfy an analogous infinite set of sum-rules.
In this case, the sum-rules can be solved, and there is
a large body of work successfully
comparing the solutions to numerical results
from overlap and related fermions (see, for example, the
review in Ref.~\cite{Damgaard}). 
It would be of considerable interest if the solutions
could be extended to the sum-rules discussed here.

One might wonder what can be learned about the other
scenario predicted by \chipt---that involving a first-order
transition. The answer appears to be very little.
In this case $\rho(\lambda)$ always has a gap, there
is no SSB, there are no massless Goldstone pions, and
thus no $\epsilon$-regime.
One expects from \chipt\ or from the
p.d.f. analysis that $\langle C_a\rangle= \langle C_a^2\rangle=0$ 
for both $a=0,3$.
The consistency of these results with
the expressions in terms of eigenvalues is almost trivial,
because with a gap there are no infrared divergences.

One might also wonder what happens to the present
analysis in the continuum limit.
In particular, how does it connect 
with that of Leutwyler and Smilga?
The short answer is that there is no direct connection, since
the limits $a\to0$ and ${\Nsite}\to\infty$ 
do not commute when in the Aoki-phase.
If one takes ${\Nsite}\to\infty$ first, as I have throughout,
one is always in the $p$-regime for the neutral pion, 
so that the vacuum manifold is $SU(2)/U(1)$,
while, with $a\to 0$ first
(and at the same time sending $m\to0$ so as to remain in the Aoki-phase),
the manifold is $SU(2)$.
Another way of seeing the difference is to note that, if ${\Nsite}\to\infty$
before $a\to0$, then would-be zero modes are completely buried
in the continuum of near-zero modes, while if $a\to0$ first then the
would-be zero modes lie below the continuum of near-zero modes.

One can, nevertheless, ask what happens if $a\to0$ first.
One still expects $\langle C_0^2\rangle_\epsilon=0$
(from \chipt\ or from the properties of p.d.f.s),
and must understand this result.
The answer turns out to be that zero-mode contributions to
both $\langle C_0^2\rangle_{{\rm disc},\epsilon}$
and $\langle C_0^2\rangle_{{\rm conn},\epsilon}$ conspire to
cancel the contribution from the continuum of near-zero modes
to $\langle C_0^2\rangle_{{\rm conn},\epsilon}$.
The cancellation occurs as long as one of the sum-rules of
Leutwyler and Smilga holds. 
Thus one could obtain
this sum-rule by enforcing $\langle C_0^2\rangle_\epsilon=0$.
In fact, it is possible to obtain the whole set of higher-order
sum-rules by enforcing the \chipt\ relations between condensates
of higher powers. This shows the 
close relation between the methods used here
and those of Ref.~\cite{LeutSm}.

\section*{Acknowledgments}

I am grateful to 
Vicente Azcoiti, Barak Bringolz, 
Giuseppe Di Carlo, Maarten Golterman, Yigal Shamir and
Alejandro Vaquero for 
comments and discussions.

\appendix
\section{Spectral asymmetry and topological susceptibility at $a\ne0$}
\label{app:top}
This appendix provides a more detailed argument for the result  
(\ref{eq:nasym})
used in the main text for the spectral asymmetry of $H_W$.
The argument consists of two parts.
The first aims to justify the approximate proportionality
\begin{equation}
\langle |n_{\rm asym}(m_0)|^2\rangle \propto 
\langle Q_{\rm top}(m_0)^2\rangle \equiv V \chi_t(m_0)
\,,
\label{eq:nasymproptoQtop}
\end{equation}
where $\chi_t$ is the topological susceptibility,
and the arguments $m_0$ indicate that all quantities
are evaluated in the Aoki-phase.
This proportionality is for fixed $m_0$,
and thus concerns the dependence on $a$ and $V$.
The second part of the argument gives a derivation of the parametric
form of the $\chi_t$ in the Aoki-phase using 
``Wilson fermion \chipt'' (W\chipt)~\cite{ShSi,WChpt}.
Combining these two parts leads to the desired result.

From the work of Refs.~\cite{Iwasaki,SmitVink,NN95,Edwards} 
we have a fairly clear picture of how the spectral asymmetry occurs.
Given a configuration, one considers the spectrum of
low-lying eigenvalues of the ``valence'' Hermitian
Wilson-Dirac operator, $H_W(m_V)$, in which
$m_0$ is replaced by the bare valence quark-mass $m_V$.
One  studies this spectrum as a function of $m_V$, starting at
positive values and decreasing to $m_V\approx -1$.
Based on the numerical results of Ref.~\cite{Edwards}, and
subsequent theoretical work~\cite{GS,quenchedAoki}, I assume
that there is a ``valence Aoki-phase'' for a region of negative $m_V$.
For positive $m_V$ the spectral asymmetry vanishes,
but, once $m_V$ is negative and one enters the ``supercritical region'', 
eigenvalues can cross zero.
This means that, by the time one reaches the Aoki-phase,
some spectral asymmetry can have built up,
and numerical results indicate that this in fact happens~\cite{Edwards}.
The asymmetry increases as one moves through the Aoki-phase, and
becomes almost independent of $m_V$ shortly after leaving the Aoki-phase.
This approximate independence continues down to $m_V \approx -1$, 
and is due to a cancellation of modes crossing
in both directions rather than an absence of crossings.
The resulting spectral asymmetry for $m_V\approx -1$
is known to provide a robust definition of the topological charge
of the configuration~\cite{NN95}.

From the results of Ref.~\cite{Edwards}, 
this behavior appears to hold for a wide variety of ensembles,
both quenched and unquenched.
I assume here that it also holds for unquenched ensembles 
in which $m_0$ is in range leading to a dynamical Aoki-phase.
If so, then, for a given $m_0$, $n_{\rm asym}$ will
approximately track $Q_{\rm top}$ as $a$ and $V$ are
varied, leading to the result (\ref{eq:nasymproptoQtop}).
Note that 
the proportionality constant between these two quantities will
depend on $m_0$, i.e. on where one lies in the Aoki-phase.
Indeed, the behavior described in the previous paragraph implies
that $n_{\rm asym}/Q_{\rm top}$ is small near the upper boundary
of the Aoki-phase (larger $m_0$) and close to unity at the
lower boundary.

Further justification for the assumed proportionality comes
from considering the physical extent of the eigenmodes at zero-crossing.
According to Ref.~\cite{Edwards}, this 
typically decreases as $m_V$ is decreased.
Crudely speaking, one can think of the crossings which occur
before entering and within the Aoki-phase as corresponding to ``lumps''
of topological charge that extend over many lattice spacings,
and survive the continuum limit.
By contrast, those that
occur after traversing the Aoki-phase are mostly lumps of size $\sim a$.
Thus, if $m_V$ is within the Aoki-phase,
the spectral asymmetry gives 
an approximate measure of that part of $Q_{\rm top}$
resulting from lumps of greater than some minimal size.
Since the topological susceptibility in the continuum limit
(appropriately regularized~\cite{regtopcharge})
is determined by lumps of size $\sim 1/\Lambda_{\rm QCD}$, 
it is plausible that 
this ``truncated'' or ``coarse-grained'' topological charge
should lead to a susceptibility proportional to the exact
result, and thus to \Eq{eq:nasymproptoQtop}.

\bigskip

I now move to the second part of the argument. 
Accepting (\ref{eq:nasymproptoQtop}),
the next task is to determine the expected dependence
of $\chi_t$ in the Aoki-phase on $a$ (and possibly $V$). 
Given some assumptions, this can be done using W\chipt,
generalizing the standard continuum \chipt\ analysis~\cite{DiVV,LeutSm}.
As a byproduct, the form of the dependence of $\chi_t$
on $m_0$ will also be obtained,
but this does not carry over to $\langle |n_{\rm asym}(m_V)|^2\rangle$
because the proportionality constant
in (\ref{eq:nasymproptoQtop}) depends on $m_0$.

In the continuum analysis, one introduces the $\theta F\widetilde F$
term into the action, rotates it into the quark mass matrix using
an anomalous singlet axial rotation, and then evaluates 
$Z(\theta)$ using \chipt. At leading order, and in the
$p$-regime at large volume, 
$Z(\theta)=\exp[-V {\cal V}^\chi_{\rm min}(\theta)]$,
where ${\cal V}^\chi_{\rm min}$ is the ($\theta$-dependent)
minimum of the potential in the chiral Lagrangian.
Then one has
\begin{equation}
\chi_t = \lim_{V\to\infty} -\frac1V 
\frac{\partial^2 \ln Z(\theta)}{(\partial \theta)^2}\bigg|_{\theta=0}
=
\frac{\partial^2 {\cal V}^\chi_{\rm min}(\theta)}
     {(\partial \theta)^2}\bigg|_{\theta=0}
\,.
\label{eq:chitdef}
\end{equation}
To generalize this to include discretization effects, one
must certainly include $O(a^2)$ corrections to the potential,
for these contribute to the vacuum energy at leading order in W\chipt.
It is less clear, however, how to introduce $\theta$ into W\chipt.
One might consider adding a bare $\theta F\widetilde F$ term to
the lattice action, mapping this to the Symanzik action, and then
into the chiral Lagrangian---i.e. following the standard steps in
W\chipt. Such a bare operator will, however, mix with lower 
($\bar\psi\gamma_5\psi$) and higher 
($\bar \psi \gamma_5 \sigma_{\mu\nu} F_{\mu\nu} \psi$)
dimension operators, leading respectively to $O(1/a)$ effects that
need to be subtracted non-perturbatively~\cite{GuadSim} 
and $O(a)$ discretization errors.

I think, however, 
that these complications do not occur here because
the asymmetry that appears in \Eq{eq:nasymproptoQtop}
is proportional to a
$Q_{\rm top}$ that is regulated in the ultraviolet.
This precludes mixing with lower-dimensional operators.
In other words, the definition 
$Q_{\rm top}=n_{\rm asym}(m_V\!\approx\! -1)$
gives a result that requires no subtractions.
$Q_{\rm top}$ will, however, have discretization errors.
For example, using
an improved Wilson-Dirac operator in the valence
$H_W$ would lead to an different assignment of topological
charge on some configurations.
What I assume here is that this is an $O(a^2)$ error rather
than an $O(a)$ one, because one is, in effect, using valence
overlap fermions to measure $Q_{\rm top}$, and overlap
fermions have only $O(a^2)$ errors.

What this discussion leads to is the assumption
that, in order to calculate $\chi_t$,
the appropriate potential to use in the 
Symanzik continuum effective action is
\begin{equation}
V_{\rm Symanzik} \sim \bar\psi (m + i\mu \gamma_5\tau_3)\psi
+ a i \bar\psi \sigma\cdot F \psi
+ a^2 (\bar \psi \psi)^2 + i \theta F \widetilde F + O(\theta a^2)
\,.
\label{eq:Symanzikwiththeta}
\end{equation}
Here I use a schematic notation in which all constants of $O(1)$
have been dropped, and I have only shown one
of the possible forms of the chiral-symmetry-breaking four-fermion
operators. I have also dropped terms of higher order in the
chiral counting $m\sim a^2$.
In words, my assumption is that the lattice $Q_{\rm top}$
matches onto $\int F\widetilde F + O(a^2)$ in (\ref{eq:Symanzikwiththeta}).
The $O(\theta a^2)$ terms will give rise to mass independent
$O(a^2)$ corrections to $\chi_t$, which I will drop for
now but restore at the end.

Given \Eq{eq:Symanzikwiththeta} [minus the $O(\theta a^2)$ terms]
the remainder of the analysis is straightforward.
One first rotates 
$\theta$ into the fermionic terms, with each (scalar or pseudoscalar)
bilinear picking up a factor of $\exp(i\theta \gamma_5/2)$.\footnote{%
It is possible that
one can avoid the extended discussion given above 
by starting with $\theta$ inserted into the quark mass matrix
of the lattice theory, following the work of Ref.~\cite{StSe}.
I have not, however, pursued to resolution
the issues of renormalization that arise
in this approach.}
One then matches to the chiral effective theory,
obtaining the same form as for $\theta=0$ except for the substitution
$\Sigma\to e^{-i\tilde\theta}\Sigma$, where $\tilde\theta=\theta/2$.
Thus the potential becomes
\begin{eqnarray}
{{\cal V}^\chi}(\Sigma) &=& - \frac{2 B_0 f^2}{4} 
\Tr\left(e^{i\tilde\theta}M\Sigma^\dagger +
         e^{-i\tilde\theta}M^\dagger\Sigma\right)
+ \frac{c_2 }{16} 
\left[\Tr(e^{-i\tilde\theta}\Sigma+e^{i\tilde\theta}\Sigma^\dagger)\right]^2
\nonumber\\
&&
+ \frac{c'_2}{16} 
\left\{
2\Tr\left[
(e^{-i\tilde\theta}\Sigma+e^{i\tilde\theta}\Sigma^\dagger)^2
\right]
-\left[\Tr(e^{-i\tilde\theta}\Sigma+e^{i\tilde\theta}\Sigma^\dagger)\right]^2
\right\}
\,.
\end{eqnarray}
Here I have absorbed the $O(a)$ term into the mass matrix
$M=m+i\mu\tau_3$ in the usual way, which remains possible 
even when $\theta\ne 0$.
Compared to \Eq{eq:Vchi} in the main text, 
there is an additional term, that proportional to 
$c'_2\sim a^2\Lambda_{\rm QCD}^6$.
This term vanishes when $\theta=0$ due to the
properties of $SU(2)$ matrices, but is non-vanishing 
when $\theta\ne 0$. 
Recall also that $2 f^2 B_0 = c_1/m$.

To determine $\chi_t$ using (\ref{eq:chitdef}) one must minimize
${\cal V}^\chi$ with respect to variations in $\Sigma$. 
This assumes $\mu$ is kept non-zero while $V\to\infty$ so
that we are in the $p$-regime.
Inserting 
$\Sigma = \exp(i\theta_0 \hat n\cdot \vec\tau)$ one finds,
up to an irrelevant constant, that
\begin{equation}
{\cal V} = - c_1 \cos\theta_0 \cos\tilde\theta
           + c_1 (\mu/m) \sin\theta_0 n_3 \cos\tilde\theta
           + c_2 \cos^2\theta_0 \cos^2\tilde\theta
           + c'_2 (1-\cos^2\theta_0)(1 - \cos^2\tilde\theta)
\,.
\end{equation}
Minimizing with respect to $\theta_0$, and considering
only infinitesimal $\mu$, leads to
\begin{eqnarray}
\cos\theta_{0, {\rm min}} &=& \frac{c_1 \cos\tilde\theta}
 {2(c_2 \cos^2\tilde\theta - c'_2 \sin^2\tilde\theta)}
\,,
\\
{\cal V}_{\rm min}(\theta) &=&
-\frac{c_1^2}{4 c_2} + \theta^2 
\frac{c'_2 [1 - (c_1/2c_2)^2]}{4} + O(\theta^4) \,,
\end{eqnarray}
and thus
\begin{equation}
\chi_t = \frac{c'_2 [1 - (c_1/2c_2)^2]}{2} 
+O(a^2\Lambda_{\rm QCD}^6)
= \frac{c'_2 \sin^2\theta_{0,{\rm min}}}{2}
+O(a^2\Lambda_{\rm QCD}^6)
\,.
\end{equation}
One finds that the ``calculable part'' of $\chi_t$ is determined
by the new low-energy constant, $c'_2$.
I have also reinserted the $O(a^2)$ term resulting
from the $O(\theta a^2)$ contributions to \Eq{eq:Symanzikwiththeta}.
The $O(a^2)$ term is of the same size as the $c'_2$ contribution,
but does not depend on $m_0$, being simply
a discretization error in $Q_{\rm top}$
and not related to the alignment of the vacuum.
The overall conclusion is thus that $\chi_t\sim a^2 \Lambda_{\rm QCD}^6$, 
so that on a typical configuration,
$|n_{\rm asym}| \sim \sqrt{V \chi_t} \sim a \sqrt{V} \Lambda_{\rm QCD}^3$.
This is the result used in the main text.

Finally, I address what happens 
in the $\epsilon$-regime for the charged pions.
Here one must include in the calculation of $Z(\theta)$ the
integral over the direction, $\hat n$, of the condensate in
the $SU(2)/U(1)$ manifold. Following the method of Ref.~\cite{LeutSm}
I find that this adds to $V \chi_t$ a contribution
proportional to $(\mu V \Lambda_{\rm QCD}^3)^2$. In order
to be in the $\epsilon$-regime, however, this contribution must
have magnitude much smaller than unity. Thus it has no impact
on the topological susceptibility when $V\to\infty$.

\section{Alternative formulation of consistency conditions}
\label{app:alternative}

In this Appendix I describe an alternative,
and arguably more natural,
formulation of the conditions
which must be satisfied by eigenvalue distributions
in order that the standard Aoki-phase scenario remain valid.
The formulation uses the microscopic spectral density,
and the corresponding higher-order eigenvalue correlations.
For the continuum Dirac operator,
these are the quantities whose properties 
are universal in QCD-like theories, and governed by random matrix theory.
To define them one ``zooms in'' on the region of eigenvalues 
of size $\lambda \ll (a \Lambda_{\rm QCD})/N_{\rm site}$:\footnote{%
Other scaling factors are also used in the continuum literature, 
e.g. including a factor of the condensate, $\Sigma\propto \rho(0)$,
so that the eigenvalue spacing is of $O(1)$.
I prefer not to do this since $\rho(0)$ is not a constant,
but rather depends on the position in the Aoki-phase.
Note also that some authors
define $\rho(\lambda)$ without dividing by $N_{\rm site}$,
in which case an additional factor of $1/N_{\rm site}$ is
needed on the right-hand-side of eq.~(\ref{eq:rho1s}).
}
\begin{equation}
\rho_1^s(x) \equiv \rho(\lambda=\frac{x}{N_{\rm site}}) \,.
\label{eq:rho1s}
\end{equation}
Thus $\rho_1^s(x) dx$ is the average total number of eigenvalues
between $\lambda=x/N_{\rm site}$ and $(x+dx)/N_{\rm site}$.
When expressed in terms of $x$, eigenvalues on a single
configuration form a discrete set of levels,
spaced by $\sim 1/\rho(0)$.
One obtains a continuous distribution only after averaging over
configurations, and the resulting distribution has structure
(oscillations about $\rho(0)$) which is the remnant of the discrete levels. 
This is different from $\rho(\lambda)$, which,
as noted in the main text, becomes a continuous function
$\rho_U(\lambda)$ on a single configuration when $N_{\rm site}\to\infty$.
Furthermore, $\rho(\lambda)$ does not display the oscillations for
small $\lambda$ seen in $\rho_1^s(x)$, since they get averaged out
when $N_{\rm site}\to\infty$. The essential point is that
$\rho_1^s(x)$ contains extra information about the IR region
that is lost in $\rho(\lambda)$.

I assume in the following that the lattice quark mass is chosen so that
$\rho(0)\ne 0$, implying that one is in the Aoki-phase in
the standard \chipt\ description.
The eigenvalues of interest, which I call the ``IR eigenvalues'',
are those for which $\rho(\lambda)$ is approximately constant, 
i.e. for which higher-order chiral corrections are small. 
This requires $|\lambda| \ll a \Lambda_{\rm QCD}$,
which translates into a maximum $x$ of magnitude
$x_{\rm max} = c (a \Lambda_{\rm QCD}) N_{\rm site}$,
with $c\ll 1$ a positive constant.
Since $\rho_1^s(x)$ is on average a constant,
the number of eigenvalues in the range $-x_{\rm max} \le x \le x_{\rm max}$
is $\int_{-x_{\rm max}}^{x_{\rm max}} dx \rho_1^s(x)\approx 2 x_{\rm max}
\rho(0)$, and in particular is proportional to $N_{\rm site}$.

The distribution of the IR eigenvalues is encoded by $\rho_1^s(x)$,
along with higher order correlation functions, $\rho^s_k(x_1,\dots,x_k)$.
The latter are standard quantities and I use the definitions
given in Ref.~\cite{rhodef}, that, in particular, do {\em not}
include the subtraction of the ``connected part''. 
Thus, if the eigenvalues were completely uncorrelated, 
one would have, for example,
$\rho_2^s(x_1,x_2)=\rho_1^s(x_1)\rho_1^s(x_2)-\delta(x_1-x_2)\rho_1^s(x_1)$.
The second term is present because $\rho^s_2$ is a correlation between
the eigenvalues of {\em distinct} eigenvectors. To simplify 
some of the following formulae, 
I also use correlation functions in which these 
kinematic correlations are removed, e.g.
\begin{eqnarray}
\bar\rho^s_2(x_1,x_2) 
&\equiv& 
\rho_2^s(x_1,x_2)+\delta(x_1-x_2)\rho_1^s(x_1) 
\,.
\end{eqnarray}
Then for uncorrelated eigenvalues one has
$\bar\rho_k^s(x_1,\dots,x_k)= \prod_{i=1,k}\rho_1^s(x_i)$
for all $k$.
Both the $\rho_k^s$ and $\bar\rho_k^s$ are
symmetric under interchange of any two arguments.
It it also useful to note the volume dependence of
the normalization of the $\bar\rho_k^s$,
\begin{equation}
\int_{-x_{\rm max}}^{x_{\rm max}} \left(\prod_{i=1,k}  dx_i\right)
\bar\rho_k(x_1,\dots,x_k) \propto \left(N_{\rm site}\right)^k
\,,
\end{equation}
which is consistent with the expectation that $\bar\rho_k^s$ 
approaches a constant when all arguments have magnitudes much
larger than $1/\rho(0)$.

In the continuum, the corresponding correlators are
even functions of each $x_i$ separately, allowing one to work only
with $x_i\ge 0$. On the lattice, however, parity
invariance of $H_W$ implies only that the $\rho_k^s$ 
do not change when all arguments change sign simultaneously, e.g.
\begin{equation}
\rho_k^s(x_1,x_2,\dots,x_k) = \rho_k^s(-x_1,-x_2,\dots,-x_k)\,.
\end{equation}
The same holds for the $\bar\rho_k^s$.
Thus, antisymmetric parts such as
\begin{equation}
\Delta\bar\rho_2^s(x_1,x_2)\equiv
\left[\bar\rho_2^s(x_1,x_2)-\bar\rho_2^s(-x_1,x_2)\right]/2
=-\Delta\bar\rho_2^s(-x_1,x_2) 
\end{equation}
need not vanish, unlike in the continuum. 
In fact, this particular quantity must be non-vanishing
since it encodes the spectral asymmetry, which itself is
non-zero (as discussed in the main text and in Appendix~\ref{app:top}):
\begin{equation}
\langle n_{\rm asym}^2\rangle 
\equiv
\langle (N_+ - N_-)^2 \rangle
=
4 \int_0^{x_{\rm max}} dx_1 dx_2\, \Delta\bar\rho_2^s(x_1,x_2)
\,,
\label{eq:nasymfromDeltarho}
\end{equation}
(Note that the integrals here are over positive $x_i$ only.)
We do, however, learn from Eq.~(\ref{eq:nasymfromDeltarho})
that $\Delta\bar\rho_2^s$ cannot tend to a constant for
large $|x_i|$, unlike $\rho_2^s$ itself.
This is because the integration area grows as $N_{\rm site}^2$
while the integral grows only as
$\langle n_{\rm asym}^2\rangle\propto N_{\rm site}$
[from Eq.~(\ref{eq:nasymproptoQtop})].
One possible behavior is that 
$\Delta\bar\rho_2^s$ falls off for large $x_i$, and this indeed
is what is suggested by the discussion of the spectral asymmetry
in the main text. 

It is straightforward to convert the sum-rules given in the main text
into constraints on the $\rho_k^s$ and/or $\bar\rho_k^s$. 
Results are simplified by defining (following continuum usage)
\begin{equation}
\Sigma_0 = \frac{a^3}{Z_P}{\sin\theta_0 f^2 B_0}
= \pi \rho(0) \,.
\end{equation}
The sum-rule (\ref{eq:C3sumrule}) then becomes
\begin{equation}
\int_{-x_{\rm max}}^{x_{\rm max}} dx \frac{\rho_1^s(x)}{x^2}
= \frac23 \Sigma_0^2
\,.
\label{eq:C3SRA}
\end{equation}
This has the same form as the first Leutwyler-Smilga sum-rule 
for QCD~\cite{LeutSm}, 
except that $2/3$ is replaced in QCD by $1/2(2+|\nu|)$,
with $\nu$ the topological charge.
One must also exclude exact zero-modes from $\rho_1^s$ in the continuum.
Since $\rho_1^s(x)$ is symmetric, one could
restrict the integral to positive values only (and divide the
right-hand-side by two)---this
is how the sum-rules are usually expressed in the continuum.

In this and subsequent sum-rules an implicit limit of $N_{\rm site}\to\infty$
has been taken. This removes the UV contribution to the integral
(arising from the large $x$ behavior,
$\bar\rho_1^s\propto (x/N_{\rm site})^3$), so that the
dominant contribution to the integral is from the IR region.
Note also that the precise upper limit, $x_{\rm max}$, is
then irrelevant.

The next sum-rule is obtained from Eq.~(\ref{eq:C0sumrule}), and is
\begin{equation}
\int_{-x_{\rm max}}^{x_{\rm max}} dx_1 dx_2 
\frac{\bar\rho_2^s(x_1,x_2)}{x_1 x_2} = 
4 \int_{0}^{x_{\rm max}} dx_1 dx_2 
\frac{\Delta\bar\rho_2^s(x_1,x_2)}{x_1 x_2} = 
\frac{\Sigma_0^2}{3}
\,.
\label{eq:C0SRA} 
\end{equation}
The oddness of the integrand in both $x_1$ and $x_2$ picks out
the antisymmetric part $\Delta\bar\rho_2^s$. This means that
this sum-rule has no continuum analog.

In the discussion given in Sec.~\ref{sec:antiADV},
the consistency of the Aoki phase required what might appear to
be an artificial construct, namely
an antisymmetry in $\rho_U(\lambda)$,
subleading as $N_{\rm site}\to\infty$, and yet contributing 
to the sum-rules due to its IR divergence.
By contrast, the consistency conditions seem quite natural
in the present formulation.
The lattice symmetries allow a new quantity
to be present, i.e. $\Delta\bar\rho_2^s(x_1,x_2)$, a function
which one expects to remain non-vanishing in the
$N_{\rm site}\to \infty$ limit. 
Just as $\rho_1^s$ needs to satisfy a consistency condition
[eq.~\ref{eq:C3SRA}],
it seems natural that the new function should too.
Certainly from a mathematical point of view
there should be no barrier to satisfying (\ref{eq:C3SRA}), since
one has a {\em function} to play with and the only other constraint
is eq.~(\ref{eq:nasymfromDeltarho}).
One solution is the second example given in Sec.~\ref{sec:antiADV},
in which $\Delta\rho_2^s\propto 1/\sqrt{x_1 x_2}$.\footnote{%
This form can hold only for large $x_1$ and $x_2$.
In particular,
oddness in $x_1$ and $x_2$ separately implies that $\Delta\rho_2^s$
vanishes when $x_1\to0$ or $x_2\to0$, properties also needed if the
integral is to converge in the IR.}

The pattern of one sum-rule for each new function continues at higher
order. The sum rules are obtained from enforcing
\begin{equation}
\langle C_0^{2n}\rangle_\epsilon=
\langle C_3^{2n}C_0^{2n'} \rangle_\epsilon = 0
\quad {\rm and} \quad
\langle C_3^{2n}\rangle_\epsilon=
\frac1{2n+1} \left(\langle C_3\rangle_p \right)^{2n}
\end{equation}
for integer values of $n$, $n'$,
results that can be obtained from \chipt\ or using probability
distribution functions.
In particular, combining the sum-rules for $C_0^4$,
$C_0^2 C_3^2$ and $C_3^4$, one finds
\begin{eqnarray}
\frac{4}{15} \Sigma_0^4 &=&
\int_{-x_{\rm max}}^{x_{\rm max}} dx_1 dx_2 
\frac{\rho_2^s(x_1,x_2)}{x_1^2 x_2^2}
\,. \label{eq:SR3}
\\
\frac{2}{15} \Sigma^4
&=&
\int_{-x_{\rm max}}^{x_{\rm max}} dx_1 dx_2 dx_3 
\frac{\rho_3^s(x_1,x_2,x_3)}{x_1 x_2 x_3^2}
\,. \label{eq:SR4}
\\
\frac{1}{5} \Sigma^4
&=&
\int_{-x_{\rm max}}^{x_{\rm max}} dx_1 dx_2 dx_3 dx_4
\frac{\rho_4^s(x_1,x_2,x_3,x_4)}{x_1 x_2 x_3 x_4}
\,. \label{eq:SR5}
\end{eqnarray}
Note that it is simpler here to use the $\rho_k^s$
than the $\bar\rho_k^s$.
The first constraint picks out the continuum-like, symmetric part
of $\rho_2^s$, and indeed a result of similar form
holds in the continuum.
The other two constraints pick out parts of the three- and four-point
correlators that vanish in the continuum, but need not vanish on the
lattice. These are new, essentially independent, functions, and there
is no barrier that I can see to their satisfying the new sum-rules.

In principle, one can continue this procedure to arbitrarily high order,
and obtain an infinite set of sum-rules. This set is analogous to
those that hold for each value of $\nu$ in the continuum, except that
on the lattice one has the additional functions (those odd under
sign flips of a subset of the $x_i$) and corresponding additional sum-rules.
In the continuum, these sum-rules have been solved
to obtain the $\rho_k^s$, and from them the
distribution of individual eigenvalues (as reviewed in
Ref.~\cite{Damgaard}).
It is an interesting
challenge to extend this analysis to the lattice theory.

\end{document}